\documentclass[twocolumn,showpacs,amsmath,amssymb,prl]{revtex4}

\usepackage{amsmath}
\usepackage{amssymb}
\usepackage{graphicx}
\usepackage{dcolumn}
\usepackage{bm}

\def\be{\begin{eqnarray}}
\def\ee{\end{eqnarray}}
\def\ba{\begin{array}}
\def\ea{\end{array}}

\begin{document}


\title{Current direction induced rectification effect on (integer) quantized Hall plateaus}

\author{A. Siddiki} %
\address{Physics Department, Arnold Sommerfeld Center for
Theoretical Physics, and
Center for NanoScience, \\
Ludwig-Maximilans-Universit\"at, Theresienstrasse 37, 80333
Munich, Germany}


\begin{abstract}
Current polarization induced rectification of the quantized Hall
plateaus (QHPs) is studied within a Hartree type mean field
approximation for asymmetrically depleted samples. We first
investigate the existence of the current carrying incompressible
strips (ISs), by solving the self-consistent equations, and their
influence on magneto-transport (MT) properties. Next, the widths
of the ISs are examined in terms of the steepness of the confining
potential profile considering gate defined Hall bars. The
corresponding MT coefficients are calculated using a local Ohm's
law for a large fixed current and are compared for symmetric and
asymmetric depleted samples. We predict that, the extend of the
QHPs strongly depend on the current polarization, in the out of
linear response regime, when considering asymmetrically depleted
samples. Our results, concerning the extend of the QHPs depending
on the current polarization are in contrast to the ones of the
conventional theories of the integer quantized Hall effect (IQHE).
We propose certain experimental conditions to test our theoretical
predictions at high mobility, narrow samples.
\end{abstract}
\maketitle Surprisingly, open questions remain even nowadays in
the theory of the IQHE almost three decades after its
discovery~\cite{vKlitzing80:494}. When a two dimensional electron
system (2DES) is subject to a strong perpendicular magnetic $B$
field, the energy spectrum is (Landau) quantized. Due to the
gapped density of states (DOS), the measured longitudinal and Hall
resistances, $R_L$ and $R_H$, present anomalies if the electron
density ($n_{el}$) is an integer multiple of the quantized
magnetic flux density ($n_{\phi}$), such that $R_L=0$ and Hall
resistance becomes quantized, i.e. $R_H=\frac{e^2}{\nu h}$, where
filling factor $\nu(=n_{el}/n_{\phi})$ is an integer, $e$ is the
electron charge and $h$ is Planck's constant. In a first order
approximation, two main schools have emerged in giving an
explanation to the IQHE, namely the bulk~\cite{Laughlin81} and the
edge~\cite{Halperin82:2185,Buettiker88:317,Chklovskii92:4026}
pictures, which are thought to be unavoidably in contrast to each
other in answering the question `` where does the current flow
?''. Moreover, there is an ongoing debate about whether transport
at the edges occur in the
\emph{compressible}~\cite{Halperin82:2185,Chklovskii92:4026} or in
the
\emph{incompressible}~\cite{Chang90:871,Guven03:115327,siddiki2004,Akera06:}
states, which are formed as a direct consequence of Landau
quantization and their widths are determined by the Coulomb
interaction. The local probe experiments present a strong evidence
suggesting that the current is carried by the incompressible
strips~\cite{Ahlswede02:165,Yacoby04:328}. In particular,
experiments performed at the von Klitzing's group where a scanning
force microscope was used to measure the spatial distribution of
the Hall potential across the 2DES as a function of the $B$
field~\cite{Ahlswede02:165}. The observed dependence of the
potential profile on $\nu$ already suggests the dominant role of
the $e-e$ interactions, leading to finite widths of both
compressible and incompressible strips (ISs), where the current is
carried by the later. The Hall potential profiles were categorized
mainly to three types: Type I, the potential varies linearly in
position $1.6<\nu<2$, Type II non-linear spatial variation
($\nu\approx2$) and, Type III, where the potential strongly varies
at the edges, however, is constant at bulk, $2.05<\nu<2.3$. The
observations were explained within the self-consistent (SC)
Thomas-Fermi-Poisson theory of screening~\cite{Lier94:7757} plus
the local Ohm's law (LOL)~\cite{Guven03:115327}, which are the
bases of the present work to be discussed later. In a subsequent
theoretical work~\cite{siddiki2004} the effect of finite extent of
the wave functions on the incompressible strips (ISs) was
simulated by a spatial averaging of the local quantities over
quantum mechanical length scales such as the Fermi wavelength
$\lambda$ or magnetic length $l$ $(=\sqrt{\hbar/eB}$. The spatial
averaging also enabled them to relax the strict local
approximation considering MT and to lift the artifacts arising
from the Thomas-Fermi approximation (TFA). The main outcome of
this work was to show explicitly that, if there exists an IS
somewhere in the sample (which is translation invariant in the
current direction) the system is in the QH regime, i.e. the widths
of the QHPs strongly depend on the widths of the ISs. Moreover,
its predictions on the asymmetry of the QHPs with respect to the
classical Hall resistance curve, depending on the mobility and
sample width, are confirmed experimentally~\cite{josePHYSE}.

In the present work, we first present our geometry and the related
electrostatic problem, which in turn determines the width of the
ISs. Next, we investigate the effect of a large current on the
local electron density $n_{\rm el}(x)$ using the LOL. The DOS
$D(E)$ and the MT coefficients are obtained from the SC Born
approximation~\cite{Ando82:437}. The current density
$\textbf{j}(\textbf{r})$, $n_{\rm el}(x)$ and $R_H$ are compared
in the out of linear response regime (LRR) for generic and
asymmetrically depleted samples. The asymmetric distribution of
the ISs with respect to the center of the sample and its effect on
$R_H$ is utilized as an experimental test for two different
boundary conditions, thereby density profiles. At asymmetrically
depleted samples, in which the potential profile is steeper at one
side than the other, we show that the IS at the steep edge is
narrower than the one at the smoother edge, which in turn
determines the current distribution and the extend of QHPs. As a
result of the applied current a Hall potential develops within the
sample, whose slope is determined by the current direction and
amplitude. Therefore, if the Hall potential added to the
electrostatic potential has the same slope sign with the SC
potential where the narrow IS resides, this IS is enlarged and the
QHP becomes wider otherwise, becomes narrower. Such a
rectification effect due to current polarization is counter
intuitive considering the conventional theories, both bulk and the
edge, since the extend of the QHPs mainly depend on the mobility
at fixed temperature and current amplitude, not its direction.
\begin{figure}
{\centering
\includegraphics[width=1.\linewidth]{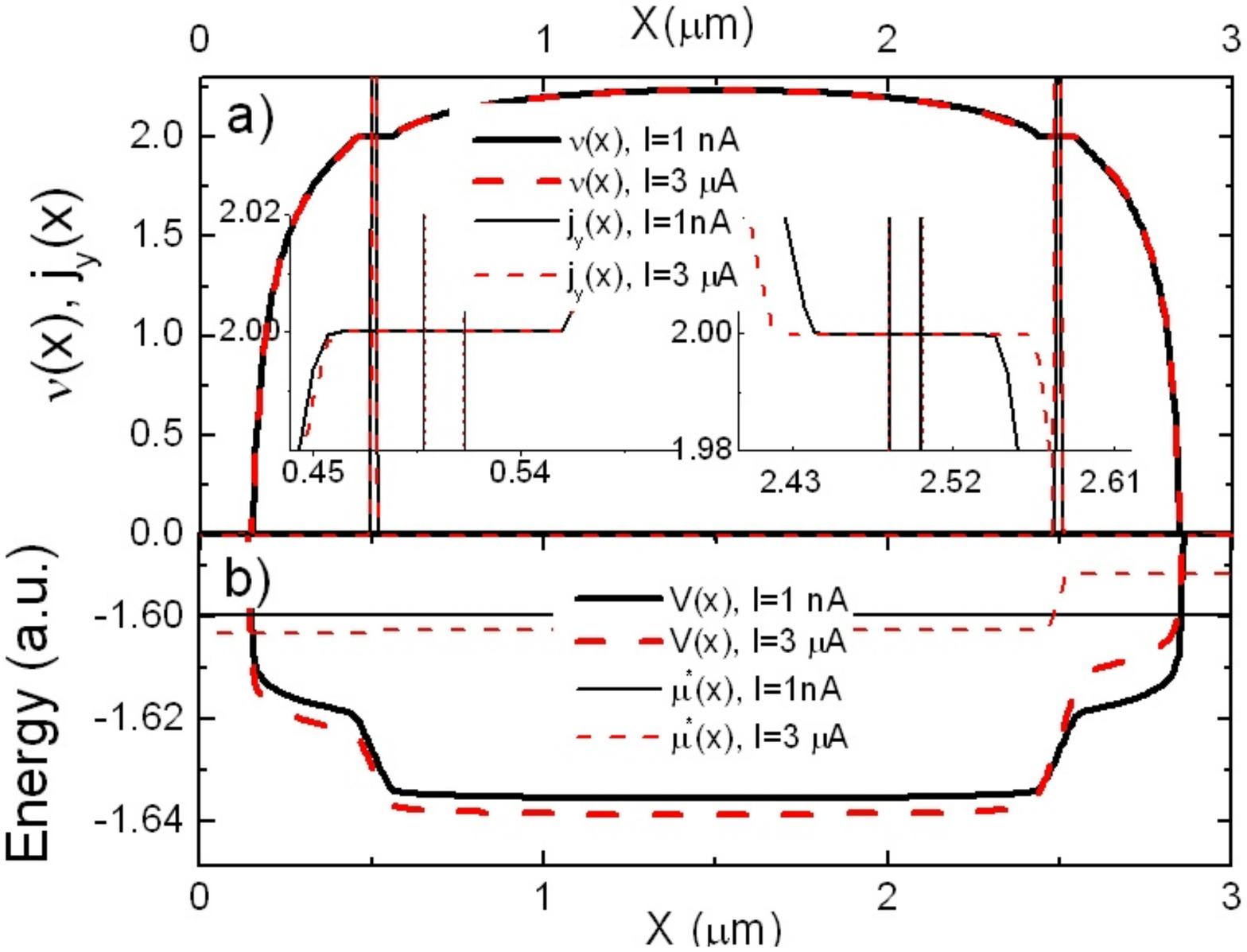}
\caption{ \label{fig:0}(Color online) (a) The spatial distribution
of $\nu(x)$ and the corresponding $j_y(x)$ considering a generic
(symmetric, i.e. $b_l=b_r=$ 150 nm, assuming $V_L=V_R=0$) 3 $\mu$m
sample. (b) The SC and electrochemical potentials under low (solid
lines) and high bias (broken lines). All calculations are done for
$\hbar(eB/m)/E_F^0$ ($=\Omega/E_F^0$)=0.94, where $E_F^0$ (=12.75
meV) is the Fermi energy at the center, i.e. $X=1.5$ $\mu$m, at
the default temperature $k_BT/\Omega=0.025$. A homogeneous donor
density of $n_0=4\times10^{11}$ cm$^{-2}$ is assumed, with the
impurity parameter $\Gamma/\Omega=0.03$.}}
\end{figure}
Here we follow the path of Ref.~\cite{siddiki2004} in describing
our 2DES considering the historical Chklovskii
geometry~\cite{Chklovskii93:12605}, i.e. a translation invariance
in $y-$ (current) direction, where donor distribution $n_0$ is
assumed to be homogeneous residing together with the electron
layer on $z=0$ plane and the 2DES is depleted from the edges by
applying $V_L$ and $V_R$ to the metallic gates on sides. In the
screening theory of the IQHE, the Coulomb interaction is included
to a spin degenerate single particle Hamiltonian, within the a
Hartree type approximation, via adding an effective mean field
potential given by \be
V_{H}(x)=\frac{2e^2}{\kappa}\int_{-d}^{d}n_{\rm el}(t)K(x,t)dt,
\label{eq:tfapot}\ee where $\kappa$ is an average dielectric
constant (=12.4 for GaAs) at the interface of the 2DES and the
kernel $K(x,t)$ preserves the boundary conditions $V(-d)=V_L$ and
$V(d)=V_R$ for the above described model. The total potential
(energy) is then \be V(x)=V_{bg}(x)+V_{G}(x)+V_{H}(x)
\label{eq:vtot} \ee where $V_{bg}(x)$ is the background potential
generated by the donors and $V_{G}(x)$ by the gates for a sample
width of $2d$. To calculate the Hartree potential one needs the
electron density distribution, which is given within TFA by \be
n_{\rm el}(x)=\int dE
D(E)\left[e^{\frac{(E-\mu(x))}{k_BT}}+1\right]^{-1},\label{eq:tfadens}\ee
where $k_B$ is the Boltzmann constant, $T$ the temperature,
$\mu(x)=\mu^{*}_{\rm eq}-V(x)$ the electrochemical potential and
$\mu^{*}_{\rm eq}$ the chemical potential at equilibrium. We
define the widths $\Gamma$ of the broadened Landau levels (LLs)
from the mobility dependent short range
broadening~\cite{Ando82:437,siddiki2004}. The SC scheme is closed
by the Eqns.~\ref{eq:vtot} and~\ref{eq:tfadens} provided that the
left and right depletion lengths, $b_l$ and $b_r$, are given. The
numerical task is now to solve these equations by iteration until
the electron density distribution remains unchanged up to a
numerical accuracy of $10^{-8}$. In the next step the local
current density $\textbf{j}(\textbf{r})$ is calculated assuming a
fixed current in $y$ direction $I=\int_{-d}^{d}j_y(x,y) dx$ via
Ohm's law \be \bf{\nabla}\mu^*(\textbf{r})/e\equiv
\textbf{E}(\textbf{r})=\hat{\rho}(\textbf{r})\textbf{j}(\textbf{r}),
\label{eq:ohms}\ee provided that the resistivity tensor
$\hat{\rho}(\textbf{r})$ is known through the
DOS~\cite{Guven03:115327,siddiki2004} and assuming a stationary
state using the local electric field $\textbf{E}(\textbf{r})$
obtained in the previous step. The translation invariance is
utilized together with the equation of continuity
$\bf{\nabla}\cdot\textbf{j}(\textbf{r})=0$ and $\bf{\nabla} \times
\textbf{E}(\textbf{r})=\textbf{0}$ to obtain \be
\begin{array}{cc}
  j_x\equiv0, & E_y(x)\equiv E_y^0, \\
  j_y(x)=E_y^0/\rho_L(x), & E_x(x)=E_y^0\rho_H(x)/\rho_L(x), \\
\end{array}\ee
where $\rho_L(x)$ and $\rho_H(x)$ are the diagonal and
off-diagonal entries of the resistivity tensor, respectively, and
the constant electric field in $y$ direction $
E_y^0=I.\left[\int_{-d}^{d}\frac{dx}{\rho_L(x)}\right]^{-1}$. Then
$\mu^*(x)$ (now position dependent) is obtained from
Eqn.~\ref{eq:ohms} by integration, up to a constant which is fixed
by $n_{\rm el}$. In our numerical scheme we start with $n_{\rm
el}(x)$ calculated without current, then calculate the current
distribution for a given fixed $I$. Next, we obtain $\mu^*(x)$
such that $n_{\rm el}$ is kept constant and start the new
iteration from the newly calculated $n_{\rm el}(x)$. This
procedure is continued until convergence is obtained.
\begin{figure}
{\centering
\includegraphics[width=1.\linewidth]{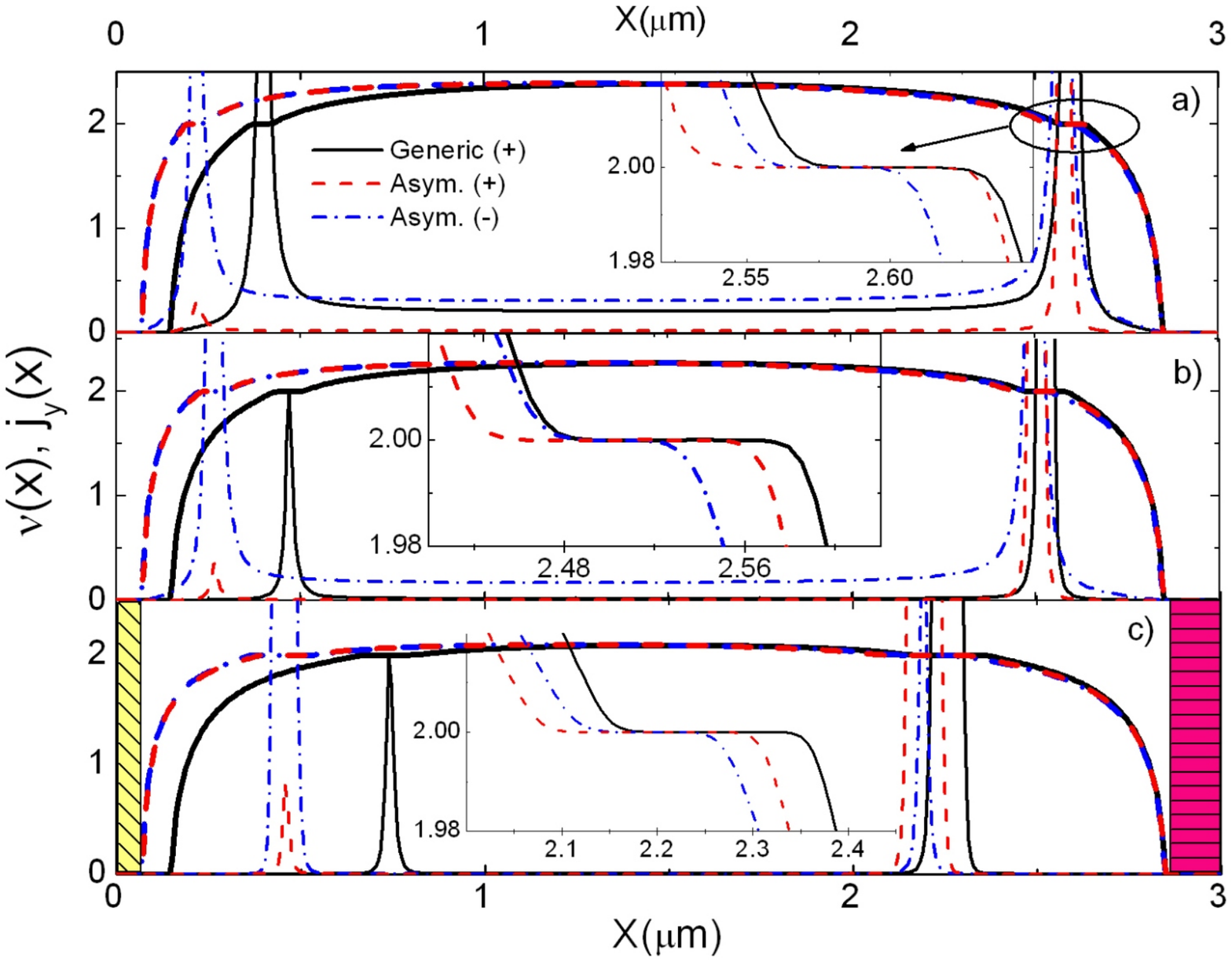}
\caption{ \label{fig:1} (Color online) The local variation of the
filling factor and the current densities for three selected $B$
values, indicated in Fig.~\ref{fig:2}, considering: (i) generic
(black solid line) and (ii) asymmetric samples, by setting $V_L=0$
and $V_R=-1.1$ V which results in $b_l=b_r/2=75$ nm. Current
polarizations are depicted by (red) broken lines for (+) and
(blue) dash-dotted lines for (-). The asymmetric depletion is
shown by the diagonal shaded region on LHS and horizontal shaded
region on RHS (c), wich results in $E_F^0=$ 13.12 meV. Insets show
the regions of ISs on the right side.}}
\end{figure} In this
paper we apply the above described calculation scheme to an
asymmetrically depleted gate defined sample. Following
Ref.~\cite{siddiki2004}, we perform a spatial averaging over
$\lambda~(\sim33$ nm) to simulate the effects of the finite extend
of the wave functions, which also lifts the local strictness of
the Ohm's law. We show that the large current induces an asymmetry
on the widths of the ISs due to the tilting of the Landau levels
as a result of self-consistency, i.e. adding the Hall potential to
the total potential and recalculating the electron density. The
amplitude of the current $I\sim 3$ $\mu$A is sufficiently large
being in the out of LRR. Our aim is, first to present this current
induced asymmetry calculated for a generic sample, which is
equally and largely depleted from both edges. Next using the side
gates, we deplete the sample asymmetrically such that the
potential on the right hand side (RHS) is smoother than that of
the left hand side (LHS). Therefore, the IS on RHS is larger even
without any current induced effect. In the last step we apply a
large negative (-) and a positive (+) DC current to the system and
investigate its effect on the density distribution and the widths
of the QHPs.
\begin{figure}
{\centering
\includegraphics[width=1.0 \linewidth]{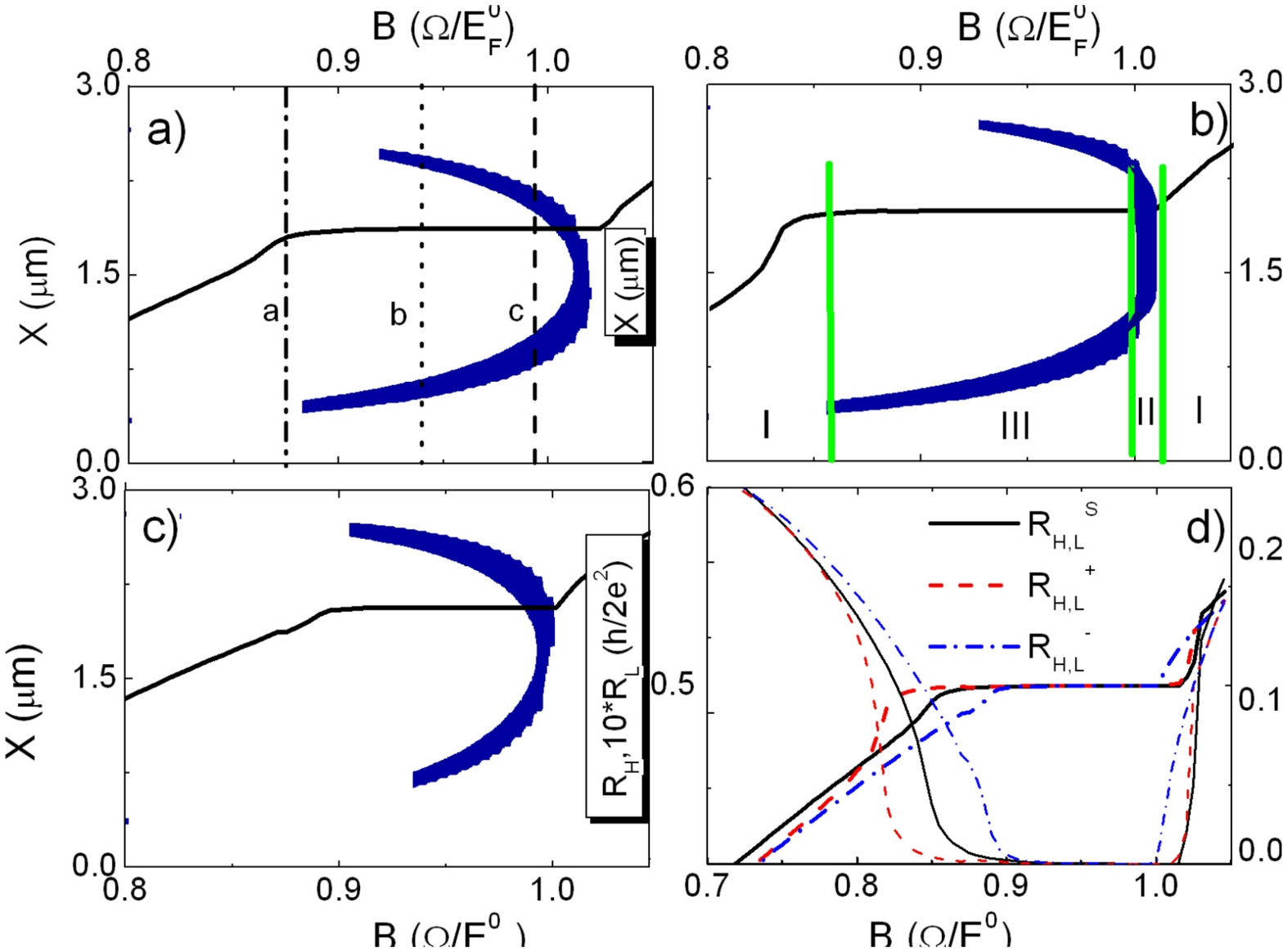}
\caption{ \label{fig:2} (Color online) The spatial distribution of
the ISs (blue regions) as a function of $B$ for generic (a) and
asymmetric samples with (+) (b) and (-) DC polarization (c). The
corresponding $R_H$ and $R_L$ (d). Vertical lines indicate the $B$
field values shown in Fig.~\ref{fig:1}.}}
\end{figure}

The applied current introduces a Hall voltage, which is added to
the SC potential, hence tilts the LLs and electrons are
redistributed accordingly. As a consequence of a (+) bias, if
there exists an IS, the spatial extend of the energy gapped region
on the RHS becomes wider, whereas shrinks on the LHS, resulting in
a wider IS on the RHS and a narrower IS on the LHS. Such a
situation is shown in Fig.\ref{fig:0}a. The local electron
distributions (or equivalently the local filling factor
$\nu(x)=2\pi l^2$) calculated at low (thick solid lines) and high
(thick broken lines) current biases is shown together with the
current density distribution (thin vertical lines) in the upper
panel for a generic sample. It is clear that, the IS on the RHS is
larger than the one on the LHS under the large current bias (cf.
the inset of Fig.~\ref{fig:0} and the current (horizontal lines)
is well confined to the ISs. Fig.~\ref{fig:0}b presents the SC
potentials (thick lines) together with the position dependent
electrochemical potentials (thin lines). We observe that at the
large bias the Hall potential tilts the LLs.  Note that, since the
compressible regions can almost perfectly screen the Hall
potential, the major effect on the $\mu^*(x)$ is observed at the
regions where an IS resides. Now, we investigate the effect of
current bias induced density asymmetry at the asymmetrically
depleted samples. In Fig.\ref{fig:1} (black) solid curves present
$\nu(x)$ of a generic sample. For the lowest $B$ value (a), no ISs
exist larger than $\lambda$ (see the inset) for the symmetric and
negatively biased asymmetric (dash-dotted blue curves) samples.
Therefore, the electron and the current densities both remain
symmetric (note that the system is completely compressible), hence
the induced Hall potential can be almost perfectly screened. The
current distribution exhibits, local spikes at the positions of
$\nu(x)\approx 2$ since, $\rho_L(x)$ assumes very small values in
the very close vicinity of $\nu(x)=2$, although no IS exists. One
can clearly see that some amount of current is still flowing from
the bulk. However, for the asymmetric sample under (+) bias
(broken red lines) the IS on the RHS is larger than $\lambda$,
hence, the current is confined within this region mainly,
meanwhile no current flows from the bulk. We observe that for this
$B$ value, the generic and asymmetric (-) samples are out of the
QHP. When increasing $B$ slightly (Fig.~\ref{fig:1}b), an IS is
now well developed on the RHS of the generic sample where most of
the current is confined to, however, due to the local minima at
the LHS mention before some current also flows form this side. The
interesting point is that now no current flows from the bulk (up
to our numerical accuracy, i.e, 10$^{-14}$ A) and the system is in
a QHP both for generic and asymmetric (+). Increasing the $B$
field furthermore, results in formation of an IS also at the
asymmetric (-) sample, where all three samples are in the plateau
regime. When the center $\nu$ becomes very close to two (not shown
here) the two ISs merge at the bulk and all the current is now
flowing from the \emph{incompressible bulk}, slightly asymmetric
with respect to the center. At the highest $B$ field strengths
shown in Fig.\ref{fig:2}, the system is out of the QHP and both
the electron and the current densities become symmetric, again. If
compared to Ahlswede experiments the sequence of the potential
types is I-I-III-III-II-I, which perfectly matches with the
findings. It is reasonable to expect that, if one starts already
with an asymmetric density profile, the asymmetry induced by the
large current will be either enhanced or suppressed depending on
the current direction. A (+) bias will tilt the LLs resulting a
high potential on the RHS, whereas a (-) bias will do the
opposite. Therefore, for the (+) bias the sequence is
I-III-III-III-II-I, which essentially means that the asymmetric
sample enters to the QHP at a lower $B$ field compared to a
generic sample, due to already existing large IS at the RHS. The
situation is rather different for the (-) bias, since the narrow
IS is on the LHS and the high bias will enlarge this IS. Hence,
there is a competition between the slope of induced Hall potential
and the confinement potential to generate a wide IS. Thus, the
asymmetric sample when (-) biased will enter to the QHP at a
higher $B$ field value compared to both (+) biased and generic
samples. The asymmetric distribution of the ISs is obvious for the
generic sample, where we only show the ISs (dark-blue regions)
wider than $\lambda$. Our findings point out that the extend of
the QHPs depend strongly on the sample asymmetry and the current
polarization. The experimental manifestation of the predicted
rectification effect requires, first of all, high mobility ($\geq
1.0\times 10^{6}$ cm${^2}$V/s) and narrow ($2d\lesssim 10$ $\mu$m)
asymmetrically depleted samples. One possible option is to define
the Hall bars similar to the ones investigated in
Ref.~\cite{josePHYSE}, at which, an asymmetry in $R_L$ is
observed. However, the effect is not pronounced to draw clear
conclusions. The main drawback of the gate defined samples relies
on the fact that, by using gates one cannot create very steep edge
potential profiles, therefore, rectification is somewhat
suppressed. Meanwhile, one can define Hall bars with steeper edge
potentials by deep etching, of course, with different etching dept
on both sides. However, it is known that etching can cause
inhomogeneities at the density profile which may become important
when considering narrow samples. A hybrid solution, i.e. one side
etch, other side gate defined, seems to be the most reasonable
solution. To obtain the extreme sharp edge on one side, it is
desirable to perform the suggested experiments on cleaved edge
overgrown (CEO) samples, where it has been shown that no ISs
reside at the sharp edge~\cite{Grayson05:016805}. The experiments
need not to be done at very low temperatures ($0.4<T<4.0$ K),
whereas the imposed current should not exceed the breakdown
current due to Joule heating~\cite{Akera06:}, which can easily be
determined by the experiments.

In conclusion, for the high mobility, narrow and asymmetric
samples we predict that, the large current either enlarges or
shrinks the QHPs depending on whether the asymmetry induced by the
current and the asymmetry caused by the edge profile coincides or
not. Based on our findings, we proposed three set of sample
structures where the effect of the current induced asymmetry and
thereby the rectification of the QHPs can be controllably
measured. As a final remark, we note that at the edge IQHE regime,
i.e. Type III, a highly non-equilibrium situation is present, due
to the competition between the enhancement of the ISs resulting
from the large current and suppression due to steep potential
profile, therefore we expect a hysteresis like behavior in this
regime both depending on the sweep rate and direction of the $B$
field and current amplitude.

Author acknowledges M. Grayson for fruitful discussions concerning
the CEO samples, R. R. Gerhardts for his valuable suggestions and
K. G\"uven for his contribution to extend our work to the
non-linear regime. This work was financially supported by NIM area
A and DIP.

\end{document}